# Diamond Nanophotonics

*Igor Aharonovich[1*] and Elke Neu[2]*

1. School of Physics and Advanced Materials, University of Technology Sydney, Ultimo, New South Wales, 2007, Australia
2. Department of Physics, University of Basel, Klingelbergstrasse 82, CH-4056 Basel, Switzerland
Email: igor.aharonovich@uts.edu.au

*The burgeoning field of nanophotonics has grown to be a major research area, primarily because of the ability to control and manipulate single quantum systems (emitters) and single photons on demand. For many years studying nanophotonic phenomena was limited to traditional semiconductors (including silicon and GaAs) and experiments were carried out predominantly at cryogenic temperatures. In the last decade, however, diamond has emerged as a new contender to study photonic phenomena at the nanoscale. Offering plethora of quantum emitters that are optically active at room temperature and ambient conditions, diamond has been exploited to demonstrate super-resolution microscopy and realize entanglement, Purcell enhancement and other quantum and classical nanophotonic effects. Elucidating the importance of diamond as a material, this review will highlight the recent achievements in the field of diamond nanophotonics, and convey a roadmap for future experiments and technological advancements.*

# 1 Introduction

Optically active impurities in diamond, so called color centers, have attracted intense research interest in recent years. The success of these emitters builds upon their unique properties that include highly stable fluorescence (photostability) of single color centers even at room temperature and feasible control of single, highly coherent spins associated with color centers in diamond[1, 2]. Over 500 different color centers in diamond are known[3]; their emission wavelengths span a spectral range from the ultraviolet to the near infrared. Furthermore, the diamond host material provides beneficial qualities like chemical inertness, biocompatibility, high transparency from the ultraviolet to infrared spectral range as well as a high mechanical strength (high Young`s modulus) and exceptionally high thermal conductivity[4-6]. Furthermore, color centers in diamond have to be considered as potential building blocks of future quantum information processing (QIP) architectures and integrated nanophotonic devices as detailed in the following.

Single quanta of light, single photons, are the fundamental constituents of QIP. First, single photons can be utilized to encode and transmit information in a way that the laws of quantum mechanics ensure a secure exchange of information (quantum cryptography, for review see e.g. [7]). Second, photons can serve as fast carriers of quantum information connecting distant nodes in so called quantum networks[8, 9]. In a quantum network, long lived quantum bits (qubits) store quantum information in so called quantum nodes and the information is exchanged via flying qubits, photons. Furthermore, in all optical quantum computing schemes[10], indistinguishable photons are proposed as building blocks of a quantum computer that could potentially outperform classical computers. Color centers in diamond are highly promising in QIP as they can serve as ultra bright, highly polarized, narrow band single photon emitters. Furthermore, similarly to quantum dots color centers possess a very short lifetime of several nanoseconds that enables fast triggering of single photon emission. Finally, the nitrogen vacancy (NV) defect supplies a coherent electron spin that can serve as a long lived qubit and second enables stable emission of single photons carrying the information of the spin state of the center[11-14]. Proof of principle experiments demonstrated QIP building blocks with NV centers like quantum entanglement between single photons and a single spin[15], creation of indistinguishable photons (two-photon-interference)[16, 17] and the manipulation and control of electron spins of color centers as well as nuclear spins coupled to color centers to store quantum information and form quantum registers (for recent review see e.g. [13].

To harness the full potential of color centers in diamond for QIP, it is crucial to efficiently collect their fluorescence and engineer their photonic properties like lifetime and emission bandwidth. These goals are most efficiently achieved using nanophotonic devices. Waveguides (e.g. nanopillars)[18] or nanoparticles (e.g. metallic nanoparticles as plasmonic antennas)[19] can increase the collection efficiency while nano or micro cavity structures can be used to engineer the emission bandwidth as well as the lifetime of the color centers[20]. Using nano-pillars, the collection efficiency has been enhanced by more than an order of magnitude compared to bulk diamond paving the way to use color centers in diamond as single photon sources[18]. Here, the collection efficiency limits the efficiency of the source and thus the bit rate. Considering the creation of indistinguishable photons, the collection efficiency is even more crucial as it is necessary to collect simultaneously emitted photons from different color centers. Additionally, the lifetime of a color center is an important figure of merit: sources with shorter lifetimes can emit single photons at a higher rate. Thus, controlling the spontaneous emission rate via cavity coupling is highly desirable. Furthermore, cavity coupling enhances the coupling of color center and light field thus cavities can be thought of as an interface between light and the color center (spin). It should be noted that nanophotonic devices need color centers placed in the correct depth and with the correct alignment with respect to the cavity fields, placing challenges on the creation of the emitters.

This review first introduces emerging color center emitters in diamond for nanophotonic applications – Section 2. Section 3 describes nanostructures that confine and guide light including waveguides, microdisc resonators and photonic crystal cavities. Fabricating such structures in diamond is challenging: on the one hand, diamond has to be structured using sophisticated etching recipes. On the other hand, thin (< 1 μm) single crystal diamond films are not straightforwardly available. Recent advances on the fabrication of such thin diamond membranes will be summarized in sections 3.2 and 3.3. Due to these challenges, "hybrid approaches" that use diamond emitters but non-diamond photonic structures as well as approaches using polycrystalline diamond films have been pursued and are introduced in section 3.1. Section 4 summarizes recent achievements using nanophotonics in diamond. The first part covers achievements related to nanophotonics and QIP – including lifetime engineering, entanglement, super-resolution in the optical far field as well as recent advances in diamond Raman lasers. We note that in this review we do not discuss the broad use of the NV center in magnetometry and sensing and refer the reader toward a recent review in the field[21].

## 2 Emerging single photon emitters for QIP

The most prominent emitter in diamond is the nitrogen vacancy (NV) center which has been employed in several proof-of-principle experiments in nanophotonics and QIP. In recent years, silicon vacancy (SiV) centers have been investigated extensively especially due to their narrow emission bandwidth. Color center based on heavier impurities (Nickel, Chromium, rare earth elements) exhibit highly promising properties, however, their fabrication remains challenging and their full photophysical properties are yet to be completely investigated. Figure 1 summarizes the emission properties of the various single photon emitters in diamond.

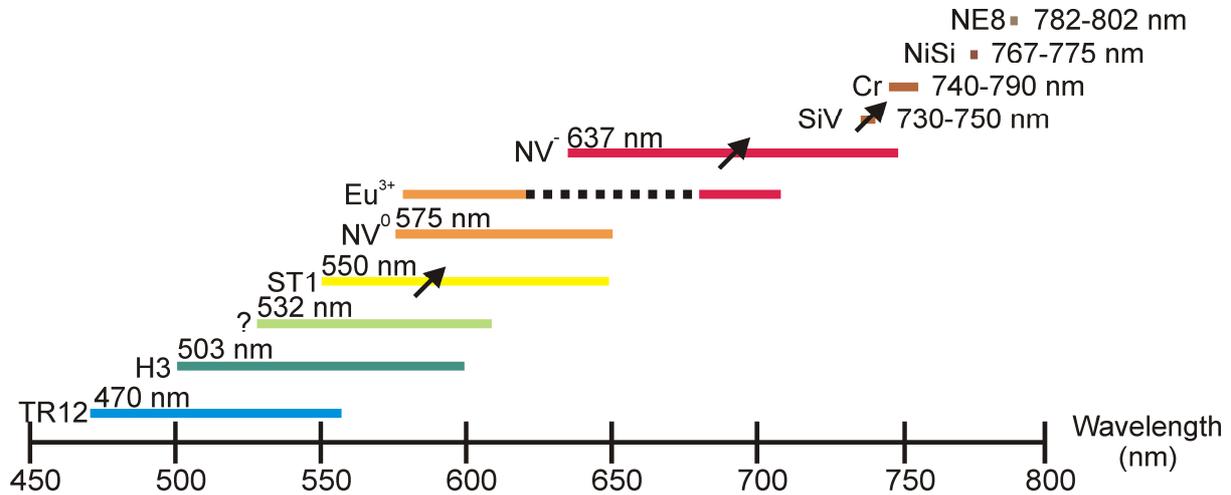

*Figure 1. Spectral map of various single emitters in diamond. For centers with emission wavelengths shorter than 730 nm, the length of the colored line represents the approximate width of the emission spectrum of the color center including phonon sidebands. The wavelength given for each center denotes the zero-phonon-line (ZPL) wavelength. Note that for the recently discovered $Eu^{3+}$ in diamond, two complex emission bands are observed both corresponding to electronic transitions and so far ensembles are observed. For centers with emission wavelengths above 730 nm, the labeled wavelength indicates the spread of the observed ZPL positions, while the colored line indicates the width of the ZPL for a single center. For these centers, the emission is mainly concentrated in the ZPL. Note that especially in the near infrared region, a multitude of unidentified color centers exists, which are not given here for clarity (see Refs.[22-25] and text for details). Centers for which spin manipulation of single centers has been shown are labeled with a black arrow.*

## 2.1 NV centers

The negatively charged nitrogen vacancy (NV) center is the first center investigated as single spin system[26] and as single photon source[2]. The immense interest into the NV center stems from its outstanding spin properties: it exhibits spin dependent fluorescence, which can be used to optically read out its spin states. Furthermore, its electron-spin has a long coherence time ($T_2$=1.8 ms in $C^{12}$ isotopically pure diamond[27]) and can be optically initialized and reliably manipulated using microwave fields. (For a very detailed recent review see Ref [28]). Thus the NV center's spin can serve as a long lived spin qubit, whereas the emitted photons can serve as flying qubits. In this case, integrating single NV centers with nanophotonic structures will enable interfacing spins and photons, paving the way to QIP.

Reliable creation of single NVs via ion implantation is crucial for their application in nanophotonics as it enables placing color centers in the desired depth in nano-structures. For

NV implantation, focused ion beams[29, 30] or nanoimplantation [31] (through a nanoscopic hole in an atomic force microscope tip), further enable a lateral placement of NV centers with a precision around 100 nm or 25 nm, respectively. Given a suitable annealing treatment, NV centers created via ion implantation have narrow, stable optical resonances[32], approaching natively occurring centers, a property highly crucial considering the coupling to high-quality cavity resonances. For MeV implantation energies, high formation efficiencies of NV centers around 50% are found[33], while for an energy of 5 keV, forming shallow NV centers (8 nm depth), efficiencies of 1% [33] up to 30% (neutral and negative charge state, $10^8$ cm$^{-2}$ dose[34]), are reported. NV centers in a defined depth are also created using the δ-doping technique, where nitrogen is introduced for a short time interval into the chemical vapor deposition (CVD) growth of diamond. Using this technique, 1-2 nm thick layers containing NV centers have been demonstrated, however, the lateral placement can´t be controlled and the efficiency of NV creation is low[35]. Very recently, NV centers in high-purity (111) oriented CVD diamonds have been shown to align almost perfectly along the [111] growth direction instead of aligning along all 4 equivalent <111> directions[36, 37]. Thus, optimal alignment of the NV`s emission dipoles with respect to photonic structures is feasible in contrast to previous work using (100) oriented diamonds[38].

Despite these major efforts in deterministic creation of NV centers, the majority of QIP experiments utilizes native NV centers and form photonic structures around a pre-characterized emitter. Furthermore, NV centers exhibit a broad emission spectrum due to strong emission into phonon-sidebands (PSBs). The purely electronic transition, the zero-phonon-line (ZPL), is found at 637 nm, the PSBs span a spectral range up to 800 nm (see figure 1). Even at 9K, only 4% of the fluorescence is emitted into the ZPL which exhibits a narrow, ideally lifetime limited linewidth (13 MHz, [39]) at low temperature; Coupling of NV centers to a photonic structure has thus either to be broadband or has to overcome the emission into the sideband by a highly efficient coupling[40]. NV centers have a lifetime of 12 ns in bulk[41], the emitted light is only partially polarized[42] and efficiency can be limited by switching to the neutral charge state[43]. There is therefore an ongoing quest to identify and characterize other emitters for photonic applications.

## 2.2 SiV centers

Negatively charged silicon vacancy (SiV) centers display a ZPL at around 740 nm. SiV centers occur only very rarely in natural diamonds, whereas they are very common impurities in synthetic CVD diamonds as a results of doping with silicon during the growth due to etching of silicon substrates, quartz reactor walls or windows in the CVD growth plasma [44-46]. Such an *in situ* creation of SiV centers has been employed to create single SiV centers in spatially isolated CVD nanodiamonds[47] or hetero-epitaxial nanoislands on iridium[48] that showed high single photon emission rates (up to 6 million counts/s[49]). However, SiV fluorescence in these material systems suffers from a large spread of emission wavelengths (ZPLs between 730 and 750 nm) supposedly due to a high level of individual stress in the nano-sized diamonds. More recently, single, bright SiV centers have been created *in situ* in high quality, low stress, single crystalline CVD diamond and several 100000 counts/s have been detected[50]. Single SiV centers were also produced via ion implantation [51, 52].

However, centers implanted deeply into natural diamond (10 MeV $Si^{2+}$, depth 2.3 μm) showed only several 1000 counts/s despite a short lifetime of 1.2 ns, indicating non-radiative decay pathways possibly due to residual damage from the ion implantation. For SiV centers, systematic studies on creation efficiency depending on the ion energy and the spatially controlled fabrication of centers and are still missing.

In contrast to the NV center, the SiV center has a very narrow emission bandwidth at room temperature: due to a weak linear electron-phonon coupling, the emission is mostly concentrated (> 70%) in the ZPL as discernible from figure 1. This property renders the SiV center especially promising as a room temperature single photon source for quantum communication [53]. Moreover, the ZPL fluorescence of the SiV center is almost fully linearly polarized at room temperature[48, 50, 54], a property highly useful in quantum cryptography where the polarization is often used to encode information. The SiV center's narrow emission bandwidth, the possibility to excite its fluorescence using red laser light or an electron beam (cathodoluminescence)[55] as well as the availability of colloidal nanodiamonds in water[56, 57], have recently triggered investigations to use nanodiamonds containing SiV centers as fluorescent labels for bioimaging of cells and living organisms[58]. Here, excitation and emission in the red or near-infrared spectral range allow for deep tissue imaging and aid in avoiding tissue autofluorescence. In this context, the recent discovery of stable SiV centers in nanodiamonds as small as 2 nm is especially interesting as it brings the size of fluorescent nanodiamonds down to the level of the size of typical dye molecules employed in biology[59]. So far these extremely small nanodiamonds with stable SiV centers have been extracted from meteoritic diamonds; for applications a route to the synthesis of such diamond materials is highly desirable.

For applications requiring a minimal bandwidth of the ZPL, cooling color centers to low temperature (typically liquid helium temperature, 4.2 K) to avoid broadening due to phonons[60] is necessary. Under such cryogenic conditions, SiV centers have demonstrated unprecedented properties of their ZPLs: single SiV centers in a high quality CVD sample showed 83% spectral overlap together with a lifetime limited linewidth of the ZPL[50]. In contrast to NV centers, where ZPL tuning via external fields was employed to achieve spectral overlap[16, 17], these results have very recently enabled the observation of two photon-interference from SiV centers without the need for spectral tuning[61].

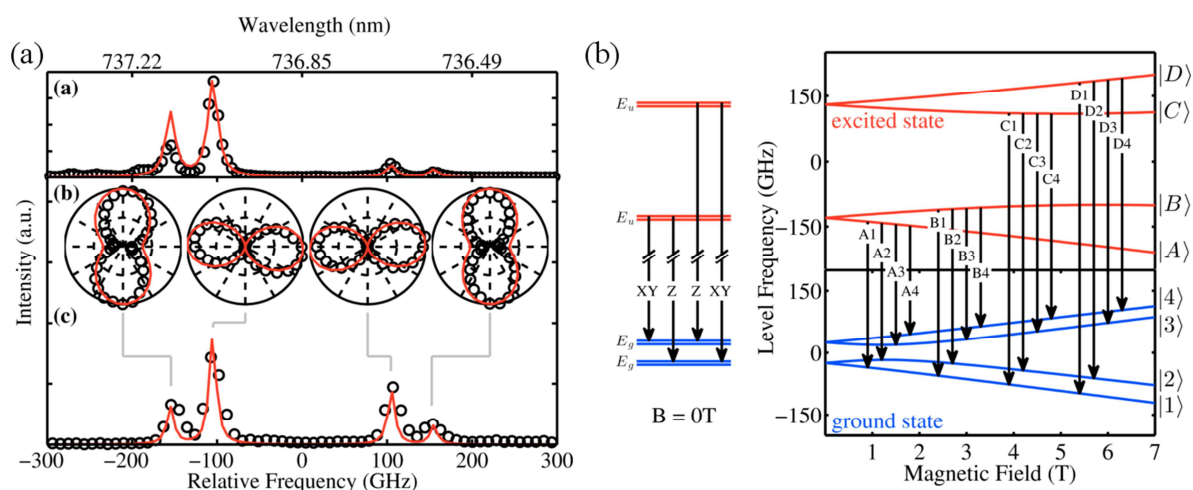

Figure 2 (A): Low temperature ZPL spectrum of the SiV ZPL. Upper spectrum for an

*ensemble of SiV centers in a high quality CVD diamond layer and lower spectrum of a single implanted SiV center under a solid immersion lens. For the spectrum of the single SiV center, the polarization is analyzed in detail (insets). (B)Level scheme of the SiV center, at zero magnetic field, ground and excited states are split into two sub-states. In a magnetic field, each of these sub-states splits into two states, leading to 16 possible transitions. For details see text. Reprinted figure with permission from Ref [51] Copyright (2014) by the American Physical Society.*

The silicon atom in the SiV center is not located on a carbon lattice site but moves along the <111> direction of the diamond lattice toward the neighboring vacancy into the so called split-vacancy site (theory see: [44, 62]). Consequently, the SiV center's symmetry class is $D_{3d}$ (compare NV center: $C_{3v}$) and the ZPL is a transition between $E_u$ and $E_g$ electronic states, which both include an orbital degeneracy. The level scheme of the SiV center is highly complex: the ZPL splits into 4 individual line components at low temperature as visible in Figure 2. The splitting of the states has been successfully modelled using a Jahn-Teller interaction and spin-orbit coupling[51]. From the polarization of these line components, it has been deduced that the SiV center has a dominant transition dipole oriented along its high symmetry axis (z-dipole) and two four times weaker dipoles in the plane perpendicular to the high symmetry axis (x-y-dipoles). Similar to the NV center, SiV centers (here neutral SiV centers) show preferential alignment (80 % out of growth plane) in diamond grown along <110> [63], indicating the possibility of aligned emission dipoles in photonic structures. The lifetime of the SiV center in high quality single crystalline diamond is 1.0 ± 0.1 ns[50], whereas in nano-sized diamonds a significant spread around 1 ns is observed[49]. Its short lifetime renders the SiV center highly suitable as a single photon source and first experiments have demonstrated quantum cryptography using the SiV as a room-temperature single photon source[53]. However, the short lifetime is most probably connected to significant non-radiative decays: Estimated quantum efficiencies for single SiVs in nanodiamonds are below 10% and in bulk diamonds, SiV centers do not reach the photon rates of NV centers in the same material[49]. However, cavity coupling offers a possibility to overcome non-radiative decays by speeding up the radiative decay[64].

One drawback of the negatively charged SiV center so far was the missing knowledge about its spin as no Electron Spin Resonance (ESR) had been clearly assigned to the center[63]. Very recent measurements of the Zeeman splitting pattern of the ZPL have identified the negatively charged SiV center as a spin ½ system[65], see Figure 2 for level scheme. Spin-selective optical transitions in a strong magnetic field have been demonstrated thus paving the way toward an all optical control of the SiV center's electron spin.

## 2.3 Additional emerging single photon emitters in diamond

Further promising single emitters are formed via incorporating heavy impurities into diamond. These centers often couple weakly to phonons thus concentrating their emission into the ZPL[66]. However, it is challenging to incorporate heavy impurities into the very tight, covalently bound diamond lattice (lattice constant 0.35 nm). Thus, a common challenge for almost all the defects mentioned in this section is that ion implantation produces them only with very low efficiency or not at all which is a drawback considering their use in nanophotonics. However, other properties might be very promising.

Nickel ions have successfully been incorporated into diamond using ion implantation[67] as well as doping during diamond growth[68]. A multitude of color centers containing nickel is known, especially in nitrogen rich high pressure high temperature (HPHT) diamond, where nickel is used as a catalyst and forms several complexes with nitrogen[69]. The only of these complexes identified as single emitter is the so called NE8 center[70-72], consisting of a nickel atom, two vacancies and four nitrogen atoms. The ZPL for single NE8 centers has been reported between 782 and 802 nm. It was either observed in natural diamond or formed by incorporating nickel during CVD growth of polycrystalline diamond or nanodiamonds. Reports on lifetime (2 and 11.2 ns) and phonon coupling are ambiguous (70% emission in ZPL in [70], 1% in [33]). Creating NE8 centers remains challenging: ion implantation and high-pressure high-temperature annealing did not yield NE8 centers[67]; recent attempts incorporating nickel into single crystal diamond during CVD only produced ensembles but no single centers[68]. Nickel-silicon-complexes (ZPL 767-775 nm)[73] and other nickel related defects, the latter formed during CVD were investigated and showed bright, narrow emission[73-75]. However, implantation efficiencies of only $10^{-6}$ render their use in nanophotonics highly challenging. Very recently, a multitude of color centers emitting between 700 and 900 nm in a CVD diamond film grown in the presence of nickel has been reported[23], demonstrating that nickel related complexes have promising emission properties but their fabrication and identification remains highly challenging.

Chromium related color centers have been reported with a narrow emission (< 11 nm) at wavelengths between 740 and 790 nm[76-81]. The first observation of these centers used nanodiamonds grown in the presence of chromium-containing sapphire substrates[82]. These defects were unique since they exhibited a two level system with no metastable (shelving state) with ultra bright, linearly polarized single photon emission. Follow up experiments employing generation of these emitters by ion implantation were not yet conclusive. In some reports, co-implantation of oxygen and chromium into type IIa CVD diamond yielded narrowband emitters[79]. Successive experiments showed that annealing (1000°C) of the same type of diamond used to produce chromium-related centers via ion implantation has led to the formation of centers with similar photo-physical properties without any implantation[22, 24]. Although the full assignment of these emitters to a particular impurity is ongoing, the first experiments toward QIP demonstrated coupling of single photons from these centers into an on-chip waveguide and interference in the on-chip devices[83].

Several other color centers in diamond have been identified as single emitters; among them centers with shorter emission wavelengths: an intrinsic center related to carbon interstitials (TR12 center, ZPL 470.5 nm[84, 85]) and the H3 center involving two nitrogen atoms and a vacancy (ZPL 503 nm[86]). Two unidentified centers where only observed once: a color center emitting at 532 nm[87] and one at 734 nm[25]. Another type of unidentified single centers (named by the authors ST1) was reported in diamond nanowires structured into high-purity HPHT diamond using an inductively coupled plasma (ICP) dry etching process [88]. The centers show a ZPL emission at 550 nm and very promising spin properties including spin dependent fluorescence like NV centers, however their origin remains unclear. The optical properties of the emitters are highlighted in Figure 1.

Very recently, the incorporation of europium into diamond has been demonstrated using an organic precursor molecule in the CVD process[89]. Rare earth elements provide very long lived spin ground states that are promising for the realization of quantum memories[90]. The possibility to realize such a system in diamond renders exploration of the incorporation of rare earth elements into diamond highly interesting. The typical luminescence of $Eu^{3+}$ was observed from nanodiamonds as well as single crystalline diamond. In the nanodiamonds, luminescence was observed from the site of individual nanodiamonds either using optical excitation or excitation using an electron beam. For the single crystalline diamond, a uniform incorporation in the grown layer was found. So far, no single emitters were detected due to a high density of the incorporated centers and their millisecond lifetime. Also, at this point fabrication of these defects using ion implantation was not successful.

# 3 Fabrication of optical resonators for the coupling of color centers in diamond

## 3.1 Hybrid approach

In the previous section, we highlighted several promising single photon emitters in diamond. The next step is coupling those emitters to optical cavities. We start our discussion by reviewing the hybrid approach. The term "hybrid approach" here summarizes techniques where diamond color centers are used as emitters but the photonic structures are fabricated from a non-diamond material. Due to the multitude of hybrid approaches to diamond nanophotonics, we here focus on some of the latest results (for further reviews on hybrid approaches see, e.g.[91]). For many applications in QIP, especially for quantum cryptography, it is highly desirable to efficiently couple single photons from diamond color centers into optical fibers to enable integration with existing telecommunication technologies. Several approaches using incorporation of nanodiamonds into the optical fiber[92], as well as placing nanodiamonds on photonic crystal fiber end facets[93] and tapered fibers have been pursued[94, 95]. Using the latter approach, 10% of the single photons from an NV center were collected using a tapered single mode fiber (diameter in taper: 260 nm). The coupling is mediated by the evanescent field of the thin, tapered region (see

Figure  (a)). The fiber mode is then adiabatically converted into the mode of the standard single mode fiber in the transition region. Highly efficient channeling of single photons from NV centers into fibers has been demonstrated using fiber based microcavities[96] in which nanodiamonds have been incorporated see

Figure  (b). Via a process termed cavity feeding, the whole phonon-broadened emission spectrum of the NV contributes to the cavity coupled emission and thus a narrowband (50 GHz), tunable single photon source has been realized.

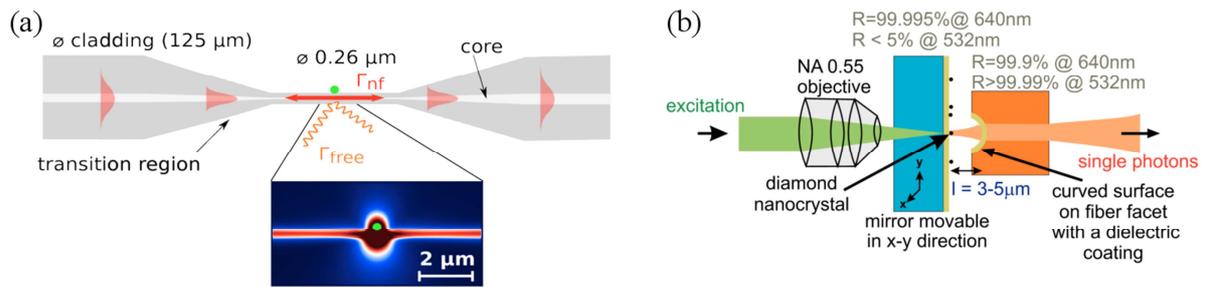

Figure 3. Fiber based hybrid approaches to diamond nanophotonics: (a) Tapered fiber coupling of nanodiamonds. A nanodiamond is placed onto the tapered region. A high coupling efficiency is feasible due to the strong evanescent field of the partially air-guided mode of the nanofiber. Via the taper, the evanescent mode and thus the single photons are converted into the mode of a standard single mode fiber. *Reproduced with permission from Ref [95].Copyright 2014 AIP Publishing LLC* (b) Fiber based microcavity for the coupling of single photons: Nanodiamonds are placed on a highly reflecting, planar mirror. The second mirror is formed in the end facet of a glass fiber, into which the photons are collected. The fiber end facet is micromachined to form a concave surface and subsequently coated with a dielectric coating. *Reprinted figure with permission from[96] Copyright (2013) by the American Physical Society.*

If the light is collected using free space optics, enhanced directivity of the color center's emission leads to higher collection efficiency. To this end, plasmonic structures can be employed. Plasmonic antennas, consisting of metal nanoparticles or nanostructures, can enhance emission directivity as well as emission rates[97]. E. g., plasmonic enhancement of NV center emission by an order of magnitude has been achieved using gold nanoparticles[19]; silver coated diamond nanowires showed a three times higher photon rate than uncoated wires[98]; sophisticated design of hybrid plasmonic cavities can yield extremely low modal volumes, and initial experiments of coupling single emitters to these cavities recently succeded[99].

Coupling NV centers to bowtie antennas has been explored to map the local effects of the antenna[100, 101]. In this respect, the single emitter was used as a probe for the local density of states of a plasmonic nanostructure and to investigate the near field interaction with graphene[102]. Recently, attaching a single nanodiamond onto an AFM cantilever enabled lifetime imaging above a metallic nanowire[103]. Therefore, coupling of quantum emitters to plasmonic structures is not only suitable for device engineering, but is also very useful to understand the performance of plasmonic structures.

Plasmonic structures can also serve as waveguides: NV centers in nanodiamonds have been successfully coupled to silver nanowires[104]. Using these silver nanowires, the emission of one NV center has been channeled into two silver nanowire-waveguides on both sides of the NV[105]. Via controlled placement of the nanowires, it has been recently shown that they can serve as a variable plasmonic beam splitter and thus distribute photons from an NV center in a plasmonic circuit (see Figure (a))[106]. However, one always has to keep in mind that the proximity of the color center to a metal surface can lead to non-radiative quenching of the emitter and that plasmonic waveguides intrinsically suffer from high losses.

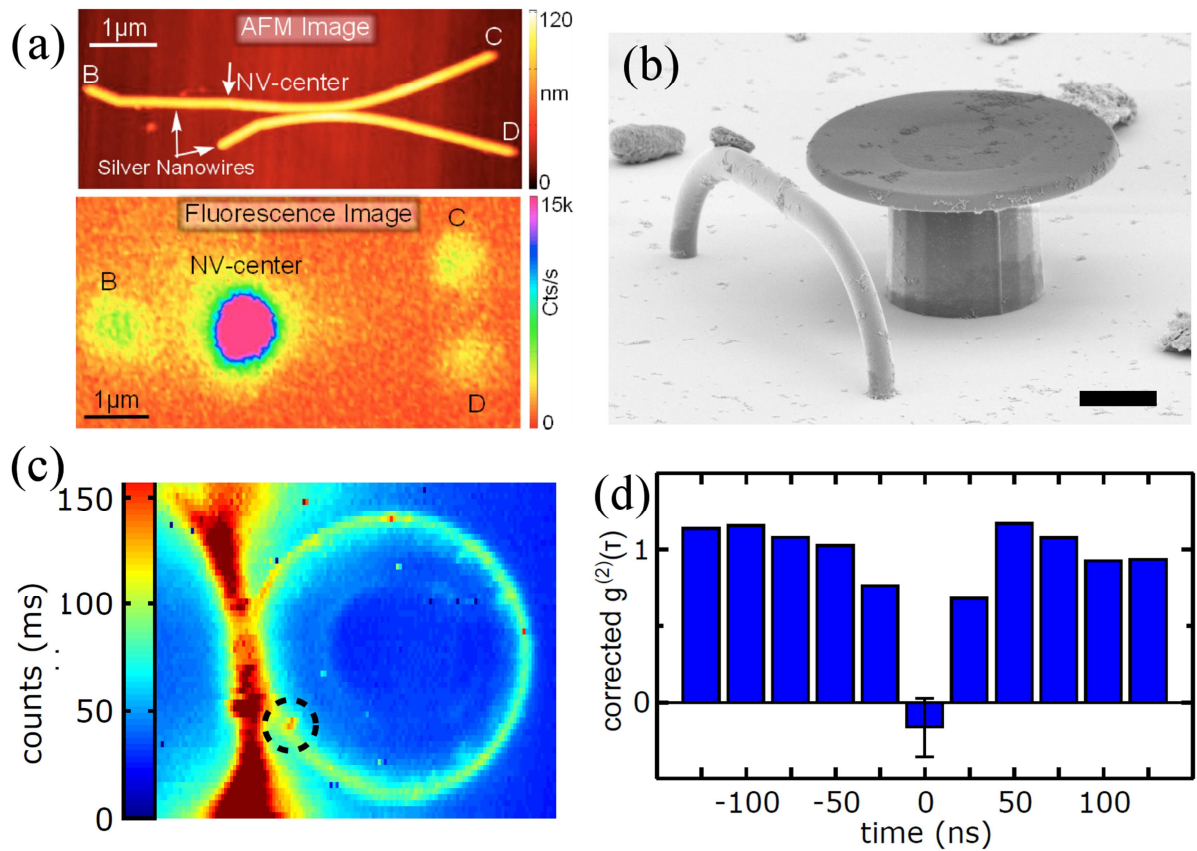

*Figure 4 Examples of hybrid approaches to diamond nanophotonics: (a) Silver Nanowires as plasmonic waveguides and beam splitter for single photons: an NV center is placed in close proximity to a silver nanowire (sample topography see upper panel). Light is guided by the nanowire as evident from the fluorescence image in the lower panel: fluorescence is not only recorded from the position of the NV center, but also from the ends of the wire (points B and C). Furthermore, the beam is split as evident by light emission from the second wire´s end (point D) Reprinted with permission from [106] (b)-(d) three-dimensional photonic structures produced via direct laser writing into photoresist containing nanodiamonds. (b) shows a scanning electron microscopy image of a disc resonator and an arc waveguide. Length of scale bars is 5 μm. (c) Fluorescence characterization of the device: the excitation laser spot is scanned across the microsdisc resonator, simultaneously; the fluorescence photons are collected at one end of the waveguide. The position of a single NV center is highlighted with a dashed circle. (d) a corresponding autocorrelation measurement is shown, the strong antibunching around zero delay confirms that a single NV center is addressed. Reprinted with permission from[107]*

Generally, the integration of single photon emitters with on chip-photonic circuits, including resonators to harvest the single photons and waveguides to route them in different directions, plays a crucial role for the development of future QIP techniques[108]. A recent approach toward integration of NV centers in nanodiamonds into three dimensional on chip structures is reported in Ref. [107]. Three dimensional photonic structures have been fabricated via direct laser writing. The technique is based on locally exposing a photoresist containing nanodiamonds via a two photon-process using focused ultrashort laser pulses. Photonic circuits consisting of microdisc resonators and waveguides have been fabricated (see Figure (b)). Coupling of the NV fluorescence to the microdisc's mode as well as the routing of single photons into a waveguide has been shown (see Figure (c)). This is the first time three dimensional photonic elements with single emitters in diamond have been realized. However, photonic structures realized in photoresist suffer from a comparably low refractive index of 1.5, thus using diamond as the photonic material (refractive index 2.4) is still desirable.

Finally, we mention here the recent progress in nanophotonic devices using polycrystalline diamond films. These can be grown on foreign substrates (e. g. silicon, $SiO_2$) on a large scale and with sub-micrometer thickness. The film's surface can be smoothed using polishing to achieve an unprecedented value of 1.6 nm root mean square (RMS)[109]. Using such thin films and under-etching of the substrate material, photonic crystals suspended in air have been fabricated and displayed Q factors of around 600 [110] in the visible range and 2800 at 1.5 µm[109]. In contrast, Ref.[111] demonstrated diamond microring resonators (Q=11000) on the $SiO_2$ substrate coupled to mm long waveguides with propagation losses of 5 dB/mm. It should be noted, that these circuits are optimal for the near infrared region (~ 1.5 µm) where the absorption and scattering from the grain boundaries is minimal. These structures can be highly useful for applications in optomechanics, however, coupling single photons from color centers is still to be proven.

## 3.2 Fabrication of nanophotonics devices from single crystal diamond membranes

### 3.2.1 "Commercial membranes" approach

To harness the full potential of diamond nanophotonic devices, structures in high-quality, high-purity diamond that host single centers with good emission and spin properties are required. Thin single crystalline diamond layers, either free standing or on a foreign substrate, indispensable for the fabrication of nanophotonic devices, are not straightforwardly available. High-quality diamond is mostly grown using CVD on a diamond substrate (homoepitaxy), thus a sacrificial layer that could be etched away to form a free-standing diamond layer is missing. Thin homoepitaxial CVD diamond layers can be fabricated by removing the substrate via laser cutting and subsequent polishing of the CVD diamond. Membranes as thin as 5 µm are available from commercial suppliers (Element Six, Sumitomo Electric and others), however, these membranes are still too thick for most nanophotonic devices and need thinning as described in the following. The starting membrane is often chosen to be thicker in the range of 20-50 µm, mostly for the ease of sample handling[112-114]. As commercially available membranes often show a wedge due to an imperfect polishing, depending on the devices that ought to be fabricated, re-polishing using a diamond mold has been used to obtain membranes with parallel surfaces. To ease handling, the thin membranes are mostly bonded to a substrate or clamped between thicker holding frames as visualized in Figure (see e.g.[113] [115]).

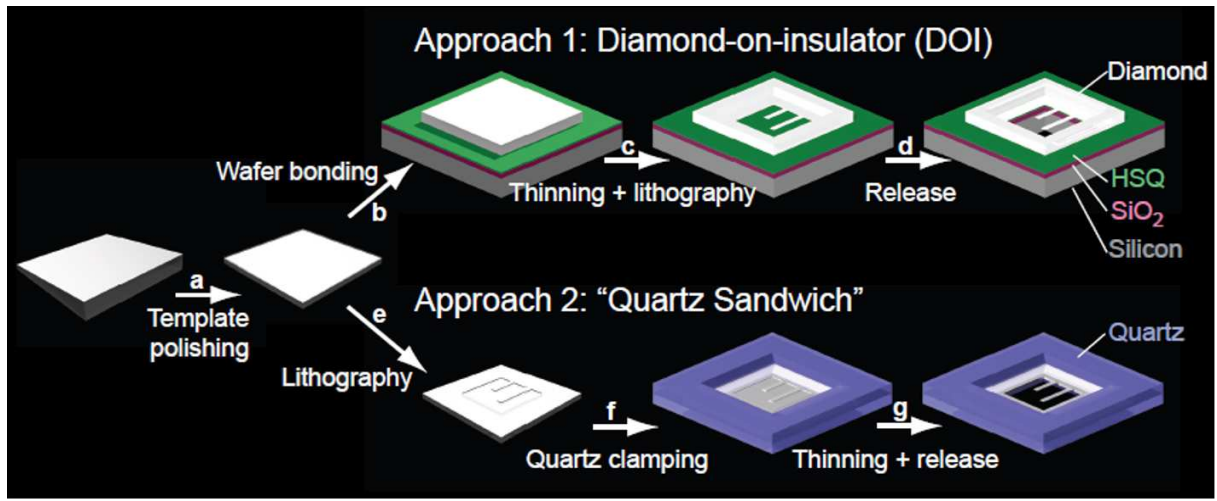

*Figure 5 Fabrication of thin diamond membranes: Commercial membranes are first re-polished to yield parallel surfaces (a). Subsequently, they are either bonded to a SiO$_2$/Si substrate using hydrogen silsesquioxane (HSQ) (b) or clamped between quartz plates with an aperture (e). In the first case, after patterning of the devices (c), backside etching of the substrate is performed to yield free standing devices. Reprinted with permission from [115]*

An alternative approach to fabricate thin diamond membranes without the need to start from already thin material has been introduced in Refs.[116, 117]. First, 200 nm stripes are defined using hydrogen silsesquioxane e-beam resist on a bulk diamond. Using this resist, trenches of about 1 μm determining the thickness of the final membrane, are etched into the diamond. Subsequently, a chromium mask is deposited and the trenches are etched to a depth of about 10 μm. As a finalizing step, the membranes are mechanically lifted out of the sample. Despite high crystalline quality of the membranes, they have so far not been employed to fabricate nanophotonic devices and given their size seem to be of limited usability for certain structures.

To thin diamond membranes, different approaches using reactive ion etching (RIE) in oxygen (O$_2$) based plasmas or argon-chlorine (Ar/Cl$_2$) based plasmas are used. The two plasma types can also be used in an alternating sequence during the etching. The use of an Ar/Cl$_2$ plasma is capable of smoothing the diamond surface[118] and conserves the surface quality of the diamond during the thinning process – a highly important aspect when making diamond photonic resonators. On the other hand, Ar/Cl$_2$ etched surfaces have been shown to contain chlorine[112] which might be detrimental considering color centers close to etched surfaces. Furthermore, the etch rate of an Ar/Cl$_2$ plasma is lower compared to an O$_2$ plasma. Thin membranes produces using O$_2$ based plasmas have been shown to retain a high crystalline quality even very close to the surfaces exposed to the plasma: Raman measurements did not show any graphite or amorphous carbon related signals and TEM measurements reveal an intact crystal lattice for the whole etched membrane. The low level of damage is attributed to the low energy of the ions in the plasma: for typical bias voltages of 250 V, ions from the plasma can only penetrate up to 0.8 nm (two monolayers) into the diamond[119].

### 3.2.2 Nanophotonic devices from commercial membranes

A multitude of devices including one and two dimensional photonic crystal cavities, ring resonators and resonators coupled to waveguides[40, 41, 114, 120-122] have been manufactured using thin diamond membranes obtained via RIE thinning. So far it has been the most promising approach to realize nanophotonic and optomechanical devices, with high quality factors as will be discussed in section 4. Figure 6 shows several examples of a microring cavities coupled to a waveguide (a) and nanobeam cavities (b)

The outstanding problem in photonic devices made using this method is the fact that the photonic crystal cavities and the microdiscs are often resting on another substrate, rather than suspended in air. To achieve full suspension, one option is using Polymethyl methacrylate (PMMA) as a sacrificial layer between the diamond and the $SiO_2$. Upon the completion of the devices fabrication, the PMMA can be selectively etched away, leaving behind a suspended photonic crystal[123]. Alternatively, a diamond membrane can be positioned on silicon, and upon patterning the diamond resonator, the silicon material from beneath the devices can be removed with an isotropic etch recipe for Si in a RIE[40, 122].

A completely different approach to the fabrication of free standing nanophotonic structures is introduced in Ref. [124]: Instead of fabricating a membrane, nanophotonic structures are etched into bulk diamond and are subsequently undercut. The undercut is achieved by an angled etching technique as illustrated in

Figure : A Faraday cage leads to anisotropic etching that occurs under an angle to the substrate's surface and thus undercuts the previously defined structures.

Figure displays an example of structures manufactured using this technique. The nanobeams displayed in

Figure show a triangular cross sections as a result of the angled etching. Unfortunately, this method is limited mostly to one dimensional photonic nanobeam cavities.

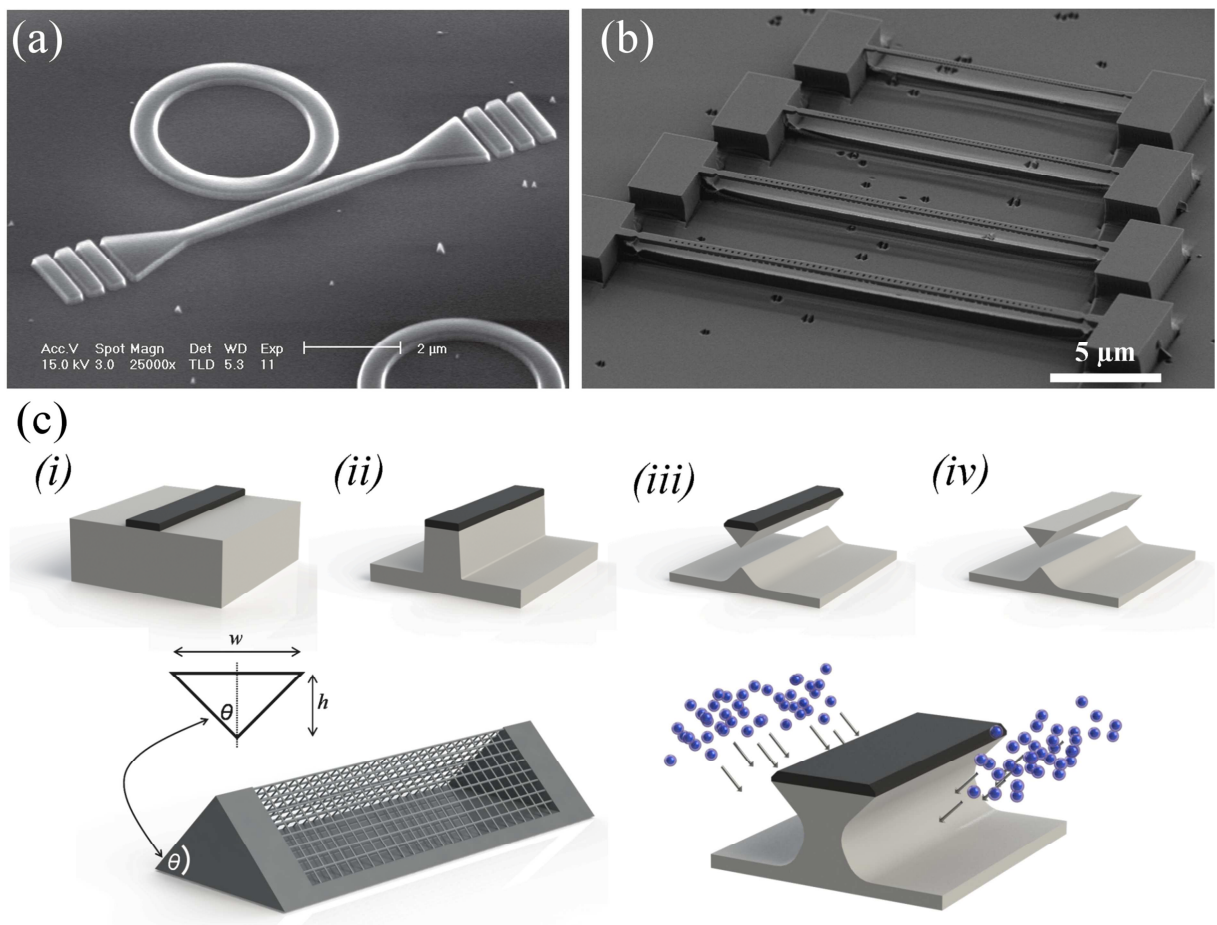

Figure 6 Realization of nanophotonic structures from single crystal diamond (a) Microring resonator coupled to a waveguide. The structure has been fabricated from an RIE thinned membrane. The device is situated on an SiO$_2$/Si substrate. Reprinted with permission from [125] (b) Nanobeam photonic crystal cavity fabricated from bulk diamond. The pattering method is illustrated in (c): In a first step, the nanobeam etch mask (i) is transferred into a bulk diamond via conventional anisotropic RIE etching (ii). Subsequently, the beam is undercut via angled RIE etching (iii) and the mask is removed (iv). In the lower panel, the angled etching process is illustrated: A triangular Faraday cage leads to an etching under an oblique angle and thus undercuts the nanobeam device. Reprinted with permission from [124]. Copyright *(2012) American Chemical Society*

### 3.2.3 Photonic devices from ion implanted membranes

Another approach toward fabrication of diamond nanophotonic devices is based upon an ion-assisted lift-off process. During this process, the lattice damage induced by the stopping of fast ions is used to generate a buried graphite layer that can be etched away to lift of a thin diamond membrane. The original idea was incepted by Parikh in the late 90s[126], however, only several years ago diamond membranes and basic optical resonators were fabricated[127]. Although rigorous analysis showed that the membranes are made out of pristine, non graphite containing diamond, no optical signature from NV centers was detected and the structures were not optically active. To understand the reasons behind the lack of optical activity, detailed Raman measurements were subsequently performed on these membranes as a function of their thickness[128]. The membranes were generated using 1 MeV He ion implantation and lift-off (membrane thickness approximately 1.7 μm), and thinned down using RIE. It was shown that the damage propagation due to implantation is extended throughout all of the membrane. Furthermore, if the region closest to the region graphitized by the ions was kept, no Raman line signature of pristine diamond was observed. The Raman signal from the far end of the membrane (i.e. the original top surface of the bulk crystal) was significantly broadened, indicating the large strain fields present in the membranes. The process of how to fabricate the membranes is shown in Figure 7a.

Although the membranes generated by ion implantation were not suitable for devices, recent work showed that they could be effectively used as templates for pristine single crystal diamond growth[129]. The process is described as follows: first, high energy ion implantation is used to generate the template diamond membranes. Then, the membranes were overgrown with a pristine diamond crystal using chemical vapor deposition (CVD), and finally, the overgrown membrane was flipped and the original "implanted" membrane was etched away using RIE. Remarkably, the overgrown CVD membranes show narrow Raman lines comparable to a pristine single crystal and were successfully employed to fabricate high quality nanophotonic resonators. In addition, ensemble spin coherence measurements of NV centers were carried out, yielding coherence times of several microseconds – comparable with high quality nanodiamonds[129]. Unlike other traditional semiconductors, where damage often propagates into the (re-)grown material (like in the case of GaN and SiC), the grown membranes exhibited clearly superior properties compared to the original damaged templates. Although detailed dislocation and crystal defect measurements are yet to be conducted, the fact that newly grown membranes were suitable for photonic devices emphasizes the overall promising path with ion implanted material. An example of a fabricated microdisk using this approach and the resulting optical properties are shown in Figure 7 (b,c).

While the technique involves multiple steps, it certainly has several advantages: first, there is no need for a laser cut of 20 – 50 μm diamond membranes and their subsequent long etch. Second, overgrowth under various conditions can be used to intentionally dope the membranes with color centers. Indeed, integration of SiV defects into optical cavities was done this way. Finally, this method can be used for growth of $\delta$ – doped layers[35]. $\delta$ – doping of nitrogen with nanometer scale depth control is enabled by introduction of nitrogen gas during a slow, high quality diamond growth. Lattice vacancies needed to form NV centers are created separately, by ex situ electron irradiation and annealing. The whole process causes minor damage in the nitrogen-doped layers compared to ion implantation techniques and has been shown to enable long spin coherence for the formed NV centers. $\delta$ – doped layers can result in a deterministic placement of the emitter in the high field region of nanophotonic cavities, therefore achieving stronger light matter interaction between the emitter and the cavity. While the first experiments of $\delta$ – doped films used a bulk crystal, growth on membranes is currently underway.

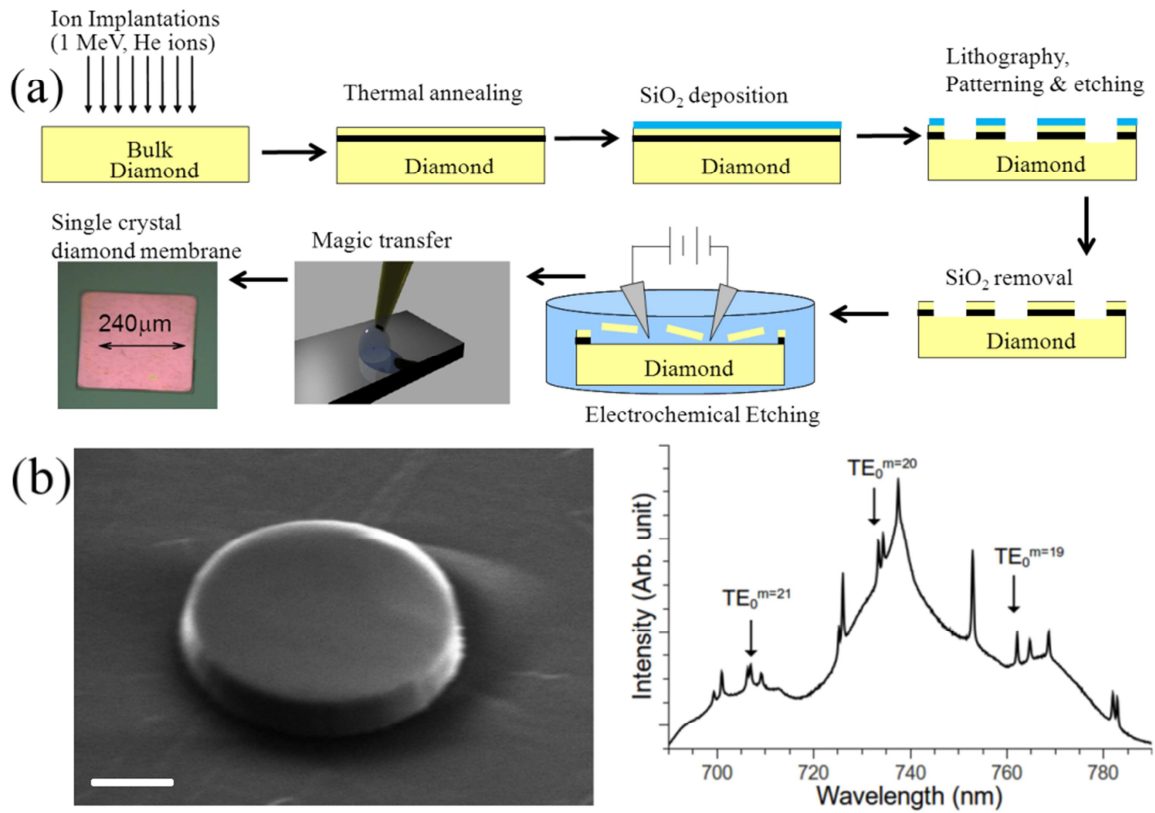

*Figure 7 (a) Generation of diamond membranes from ion implanted material. First, a CVD diamond crystal is implanted with He ions to generate a damaged region, 1.7 μm beneath the diamond surface. The sample is annealed and patterned to lithographically define the membrane size. Reactive ion etching is employed to transfer the pattern from the mask to the diamond and to subsequently etch the diamond. Electrochemical etching enables the lift-off of the membranes and a pipette is used to transfer the membrane to a substrate of choice. (b) SEM image of the microdisk made from an overgrowth process of the membrane and the corresponding and the corresponding PL spectrum from the microdisk showing high quality (~3000) factor whispering gallery modes. Images are taken from[123, 130]*

### 3.2.4  Membranes from heteroepitaxial diamond

In the previous section, we discussed fabrication of devices from homoepitaxial single crystal diamond membranes. However, these membranes are mostly limited to a size of a few mm due to the lack of larger diamond substrates. A route to overcome this limitation is diamond heteroepitaxy: A material that has a similar lattice constant than diamond and is available as large area substrate, here iridium, is used after a suitable nucleation process[131]. Although in the starting phase, the diamond has a comparably low quality as the grains from different nucleation sites merge, it can be considered single crystalline when growing to a thickness of above 10 μm. Thin heteroepitaxial membranes can be produced via etching windows into the silicon/yttria-stabilized-zirconia/iridium substrates and subsequent thinning of the diamond from the nucleation side to remove the low quality diamond[64] Although the process is slightly longer and requires sophisticated processing, the end result is promising and fully suspended membranes are possible. A clear advantage of this method is that the membranes are hold by the thick substrates which eases handling. On the other hand, since the membranes are fabricated from CVD grown diamond materials, it is likely to contain SiV defects, and is mostly suitable for nanophotonics experiments with these emitters. The fabricated photonic devices using this method are shown in Figure 8.

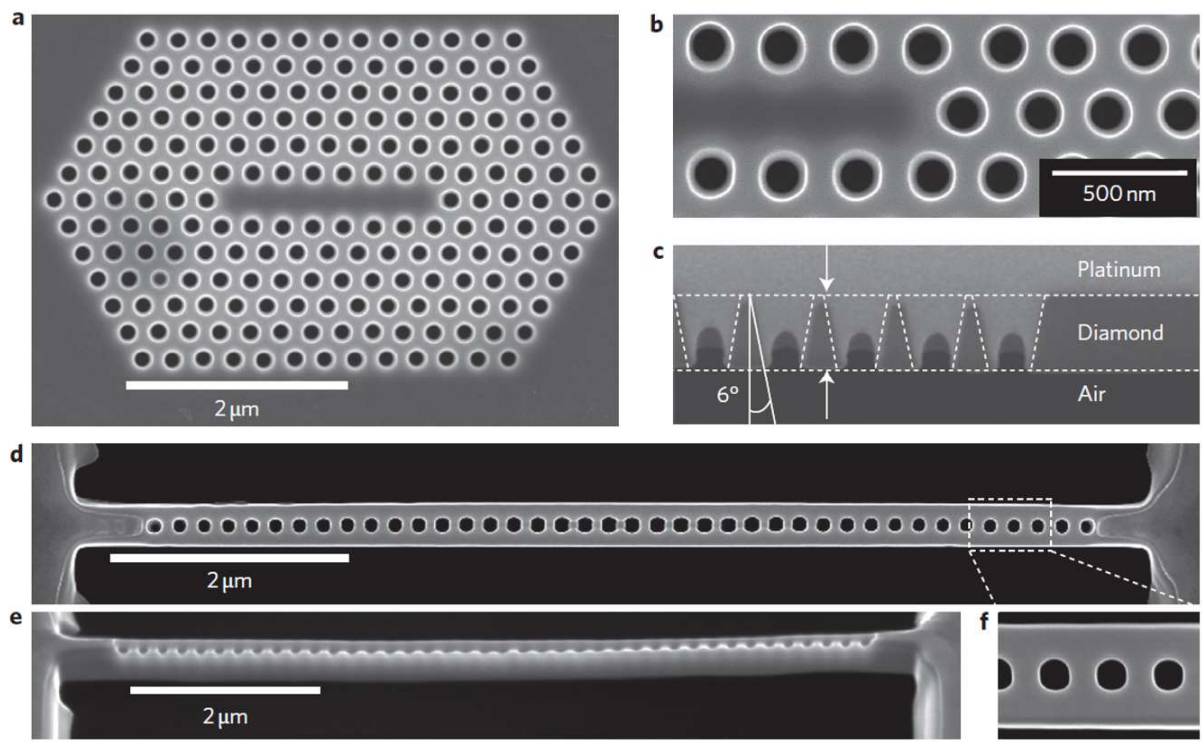

*Figure 8. SEM images of (a,b) 2D Photonic crystal caviies and (d-f) a nanobeam cavity fabricated from a heteroepitaxial diamond crystal grown on iridum substrate. Part (c) illustrates the non-vertical sidewalls of the structure due to the focussed ion beam milling. Repritned with permission from Ref[64].*

# 4 Nanophotonics experiments using color centers in diamond

We have now established the availability of various single photon emitters and different routes to fabricate photonic devices from diamond. In this section, we describe photonics experiments that have been realized with diamond devices coupled to single emitters. We will focus on both fundamental demonstrations, e.g. of the Purcell effect in nanophotonic devices, as well as practical applications including diamond Raman lasers.

## 4.1 Purcell enhancement

To realize integrated quantum photonic networks for QIP, coupling of the individual qubits to optical cavities is crucial. For the NV center in particular, coupling of the emitter to a cavity will enhance the emission into the ZPL, therefore enabling faster and brighter emission. This enhancement is vital since the ZPL contains only around 4% of the total emission of a single NV center. For semiconductor QDs (i.e. GaAs system), the dot can be pre-characterized using an AFM and the photonic crystal can be then fabricated around it. Furthermore, the excellent control over the growth enables to position the dot in the middle of the vertical direction of the photonic crystal. Such a control is yet not possible for diamond.

As mentioned in Section 2.1, creating NV centers with good properties via ion implantation is still challenging and sub 100 nm lateral placement in ion implantation is still hard to achieve. Therefore, all experiments on Purcell enhancement realized with the NV centre to date, relied on native NV defects that occur in diamond during CVD growth and the photonic structures have been manufactured "around" these centers. It is important to note, that only the radiative decay of the emitter coupled to the cavity mode is enhanced by the cavity field (Purcell effect).

The Purcell factor $F_p$ is given by the following equation[41]:

$$F = \left|\frac{\overrightarrow{E(r_i)} \cdot \overrightarrow{\mu_i}}{\|E_{max}\| \|\mu_i\|}\right|^2 \frac{1}{1 + 4Q^2\left(\frac{\lambda_i}{\lambda_{cav}} - 1\right)^2}$$

Where $\overrightarrow{\mu_i}$ is the emitter's dipole moment, $\overrightarrow{E(r_i)}$ is the local electric field, while $E_{max}$ is the maximum value of the electric field in the cavity. $\lambda_{cav}$ is the wavelength of the cavity mode while $\lambda_i$ is the emission wavelength of the color center, $V_{mode}$ is the modal volume and $n$ is the refractive index (here $n = 2.4$ for diamond). It is obvious, therefore, that to achieve optimal coupling between emitter and optical resonator, the dipole must be matched spectrally and aligned spatially with the cavity mode. This means that the emission dipole should be resonant with the cavity mode and positioned and oriented with respect to the local electric field of the cavity. In that case the Purcell enhancement factor is simplified to:

$$F_p = \frac{3}{4\pi^2}\left(\frac{\lambda}{n}\right)^3 \frac{Q}{V_{mode}}$$

It is also discernible that the highest enhancement is realized when the quality factor, $Q$, of the cavity is maximized while the modal volume, $V_{mode}$, is minimized. To date, three cavity geometries were tested to demonstrate Purcell enhancement of single NVs – microring cavity, 2D photonic crystal cavity (PCC) and a nanobeam cavity. Table 1 summarizes the important parameters of these devices and the measured Purcell effect. In all cases, tuning of the cavity mode to the NV resonance is achieved using Xe gas condensation on the resonator.

| Cavity | $V_{mode}$ | Measured $Q$ | $F_p$ | Theoretical $F_p$ | Ref |
|---|---|---|---|---|---|
| Microring resonator | ~ 17 $(\lambda/n)^3$ | Q=4300 | 12 | 19 | [41] |
| Photonic crystal cavity | ~ 0.88 $(\lambda/n)^3$ | Q=3000 | 70 | 259 | [40] |
| Nanobeam cavity | ~ 3.7 $(\lambda/n)^3$ | Q=1600 | 7 | 34 | [122] |
| Microdisk resonator | ~ 10 $(\lambda/n)^3$ | Q=3000 | - | - | [129] |

*Table 1. Comparison of performance of various photonic devices made from diamond for resonance with the NV center (~ 640 nm).*

Notably, in the case of a PCC, the lifetime modification corresponds to a spontaneous emission rate enhancement of ~ 70. This means that the cavity coupled NV center emits 70% of the photons into the ZPL. Figure 9 shows the diamond PCC and the corresponding spectroscopy measurements.

In all three cases, a significant deviation of the theoretically derived and measured enhancement is evident. There are various reasons for this discrepancy, a major one being the challenge in aligning the NV's dipole moment with the cavity field. All the optical resonators made to date, were fabricated from [100] oriented diamond, while NV centers align along the [111] direction. In addition, there are still challenges in nanofabrication of diamond using RIE and achieving perfect undercut for optical isolation.

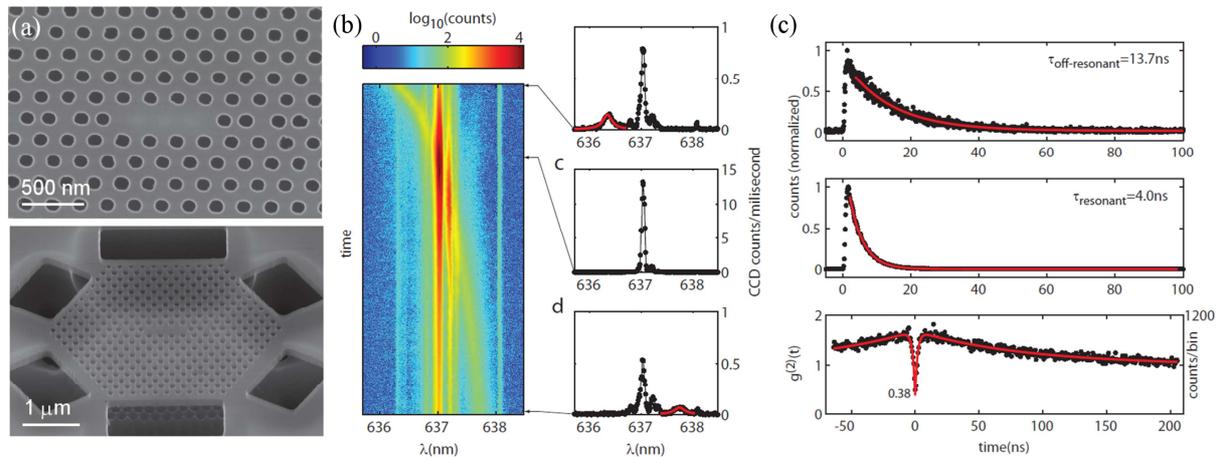

*Figure 9. Single NV centers coupled to PCC. (a) SEM image of the photonic crystal cavity fabricated from a single crystal diamond membrane. (b) Tunability of the NV ZPL into the cavity resonance. The emission is enhanced when the ZPL is overlapping with the cavity mode. (c) lifetime measurement of a single NV defect on and off cavity resonance. Lifetime reduction is observed when the emitter is on resonance with the cavity mode. Bottom panel is an antibunching measurement confirming that only a single NV center is probed. Reprinted with permission from [40]*

In addition to coupling NV centers to optical resonators, experiments on Purcell enhancement with SiV centers are underway. Initial coupling of SiVs to PCCs and microdisk cavities has been demonstrated[64, 130]. Very recently, first results on the controlled fabrication of photonic crystal cavities around a pre-characterized SiV center and the coupling of the center to the cavity have been reported[132].

### 4.2 Entanglement

The second quantum experiment to be discussed is quantum entanglement. It is seen by many as a holy grail of QIP and an unambiguous realm of quantum mechanics, refers to (in simple terms), quantum correlated objects that are remotely located. Quantum entanglement is the basis of many quantum networks, where the nodes are connected via entangled photon states[8, 133]. Entanglement has been first demonstrated using photons generated from atomic cascades, and later using spontaneous parametric down-conversion[134]. However, the strive to achieve scalability and practical devices, motivated researchers to look for a robust solid state system that can generated entangled states. To this end, a remarkable progress has been achieved with the NV center in diamond[15, 135-138].

If photons are supposed to create entanglement between distant emitters, the emitted photons must be indistinguishable. The experiment carried out by the group of Prof Hanson realized heralded entanglement between two distinct NV centers separated by more than 3 meters and located in separate liquid helium cooled cryostats[135]. Due to the properties of NV centers, several issues had to be adressed: To enhance the collected rate of single photons, solid immersion lenses were patterned around chosen NV centers. This is important, since only 4% of the emitted photons are emitted into the ZPL, and are contributing to the entanglement experiment. To suppress detection of laser photons, a cross-polarized excitation-detection scheme and a gated detection that exploits the difference between the length of the resonant excitation laser pulse (2 ns) and the excited state lifetime (12 ns) of the NV were employed. A weak magnetic field is applied to split the $m_s=\pm 1$ sublevels, and a dc electric field is used to spectrally tune the emission of two NVs (NVA and NVB) into overlap with each other via the Stark-effect. Charge state fluctuations of single NV centers reduce the probability to achieve entanglement. To minimize photo-ionization, a green laser pulse re-pumped the NV center into the desired charge state.

The entanglement protocol and the results are shown in Fig. 10 and are based on the proposal from Barrett and Kok[139]. Both NVs are first prepared in a superposition $1/\sqrt{2}(|\uparrow\rangle+|\downarrow\rangle)$ of the $m_s=\pm 1$ spin states. Then, each center is resonantly excited to an excited state with same spin projection. Spontaneous emission locally entangles the spin projection (qubit) and the photon number. The two photon modes, A and B, are directed to the two input ports of a beamsplitter, so that fluorescence observed in an output port of the beamsplitter could have originated from either of the NV centres. If the photons emitted by the two NVs are indistinguishable, detection of precisely one photon on an output port thereby projects the qubits onto the entangled state $\psi=1/\sqrt{2}(|\uparrow_A\downarrow_B\rangle \pm e^{-i\varphi}|\downarrow_A\uparrow_B\rangle)$ .[135]

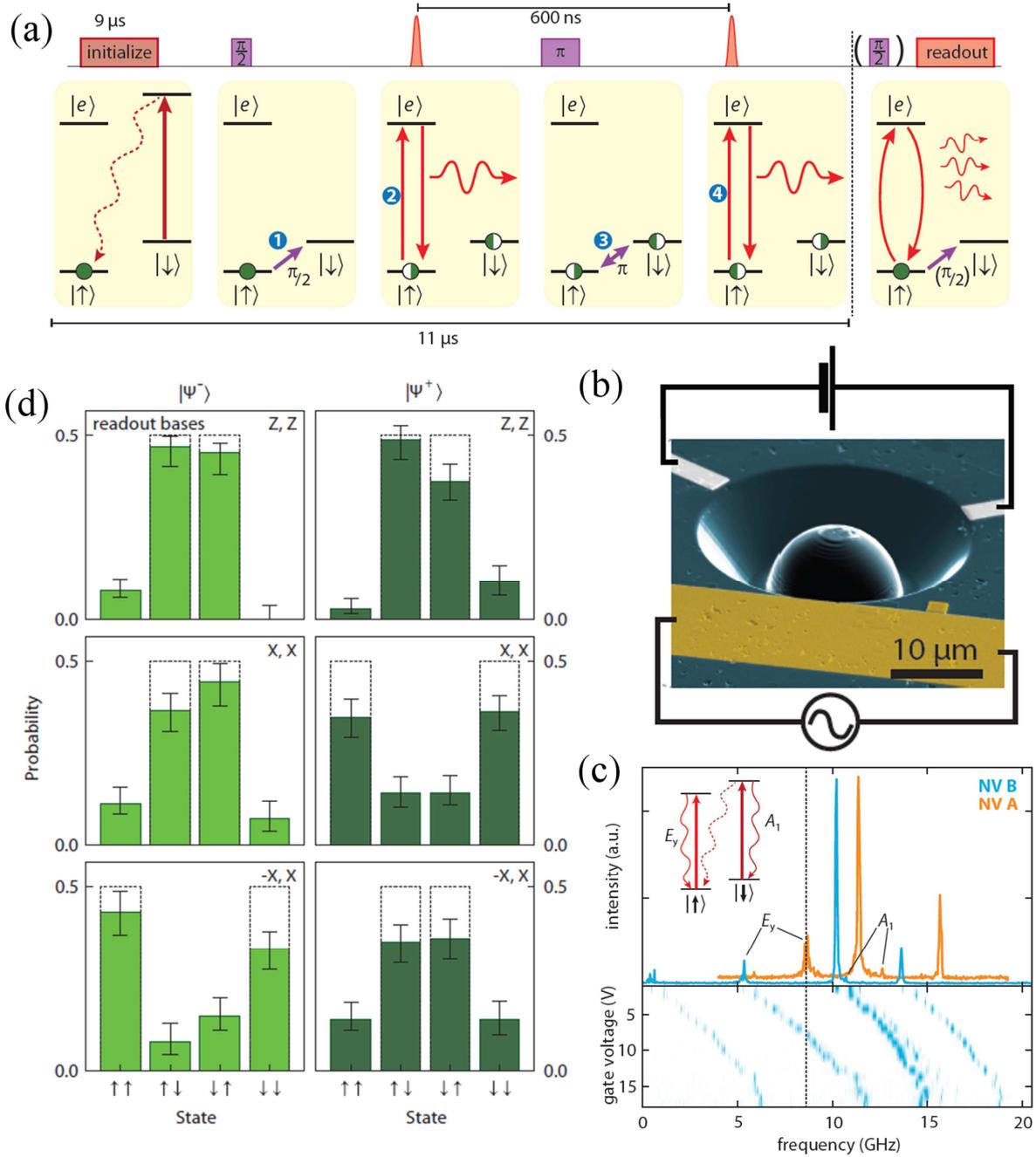

*Figure 10. Entanglement experiment with NV centers. (a) Entanglement protocol, illustrating the pulse sequence applied simultaneously to both NV centres. Both NV centres are initially prepared in a superposition $1/\sqrt{2}(|\uparrow\rangle+|\downarrow\rangle)$. A short 2 ns spin-selective resonant laser pulse creates spin–photon entanglement $1/\sqrt{2}(|\uparrow 1\rangle+|\downarrow 0\rangle)$. The photons are overlapped on the beamsplitter and detected in the two output ports. Both spins are then flipped, and the NV centres are excited a second time. The detection of one photon in each excitation round heralds the entanglement and triggers individual spin readout. (b) Solid immersion lens fabricated around the NV for the photon emission enhancement. (c) Photoluminescence excitation spectra of NV A and NV B; frequency is given relative to 470.4515 THz. Transitions are labelled according to the symmetry of their excited state. The A1 transition is used to initialize the NV centre into the $|\uparrow\rangle$ state ($m_S=0$) and the Ey transition is used for entanglement creation and single-shot readout. By applying a voltage to the gate electrodes of NV B, the Ey transitions are tuned into resonance (dashed line). (d) Verification of entanglement using spin–*

*spin correlations. Each time that entanglement is heralded the spin qubits are individually read out and their results correlated. The readout bases for NV A and NV B can be rotated by individual microwave control. The state probabilities are obtained by a maximum-likelihood estimation on the raw readout results. Error bars depict 68% confidence intervals; dashed lines indicate expected results for perfect state fidelity. Data are obtained from 739 heralding events. Image reprinted from [135]*

Before applying the protocol, both NV centres are independently prepared into the correct charge state and excited resonantly. If the preparation is successful (i.e. more than certain number of photons is emitted from NVA and NVB), the entanglement protocol is applied and repeated 300 times. A fast logic circuit monitors the photon counts and triggers single-shot qubit readout on each set-up whenever entanglement is heralded (whenever a single photon is detected). The readout projects each qubit onto the $\{|\uparrow\rangle,|\downarrow\rangle\}$ states (Z-basis) or onto the $\{|\uparrow\rangle\pm|\downarrow\rangle,|\uparrow\rangle\pm|\downarrow\rangle\}$ states (X or –X basis). By correlating the resulting single-qubit readout outcomes, the generation of the entangled states is verified[135]. The correlation results are shown in Fig 10. When both qubits are measured along Z, the states $\Psi^+$ and $\Psi^-$ display strongly anticorrelated readout results (odd parity). Furthermore, the states can be distinguished through measuring in rotated bases ({X, X},{–X, X}). For $\Psi^+$ the {X, X},({–X, X}) outcomes exhibit even (odd) parity, while the $\Psi^-$ state displays the opposite behaviour, as expected.[135]

The rate of successful entanglement between the NVs is very low, i.e. one event per 10 min, with an overall experiment duration of almost a week (!). Despite the low rate, the demonstration of entanglement between distant spin qubits is without a doubt the first significant step toward integrated quantum networks and quantum teleportation[140].

## 4.3 Super resolution microscopy

In addition to quantum applications, single defects in diamond have been exploited in realization of super resolution microscopy and sub-diffraction imaging. This chapter will highlight the recent results in this field. The unparalleled photostability of the NV center was key to realize super resolution imaging, and record fine resolutions down to several nanometers have been realized in bulk diamond. Since the original work on Stimulated Emission Depletion (STED) imaging of NV centers in diamond[141], several other methods including structured illumination and wide field super resolution imaging using spin dependent fluorescence have been developed. The latter method was explored to realize multicolour imaging using nanodiamonds[142]. Adding to the optical domain the microwave control of the emitters' spin state, the optical emission of each nanodiamond was correlated to its electron spin resonance. In other words "the colour" of each centre corresponds to a frequency in the microwave regime. This multicolour aspect of the technique enables multispectral labelling with sub-diffraction resolution, which is powerful when is utilized in biological applications. This method is shown in Figure 11a.

An additional interesting and useful aspect of the super-resolution microscopy is resolving multiple NV centers within a single nanodiamond. Some nanodiamonds contain more than a single NV center, and are often discarded for quantum applications. As reported by S Arroyo-Camejo and colleagues[143], individual NV centres were resolved within a single 100 nm nanodiamond with a resolution of ~ 30 nm. As exemplified in figure 11, individual NVs were isolated within a single nanodiamonds and their spin states could be resolved. In an independent work by M Gu and colleagues[144], the NV centers were resolved with 12 nm resolution within a single nanocrystal. In their work, M. Gu and colleagues took advantage of the stochastic photoluminescence blinking of single NV defects. The centroids of blinking NV centers were recorded at different time sequences and reconstructed, yielding a super-resolved image with the localization precision determined by the number of photons collected. Unlike single florescent dye molecules, the NV centers do not bleach, and a signal from many blinking cycles can be collected, generating a high signal to noise ratio super resolution image. The results are shown in Figure 11b.

While super resolution microscopy was employed to resolve NV centres, surprisingly, not much has been done on other emitters. One reason is the weak phonon side band that makes the choice of a STED laser complicated (i.e. for the case of SiV). However, stochastic methods that are based on blinking can certainly be used.

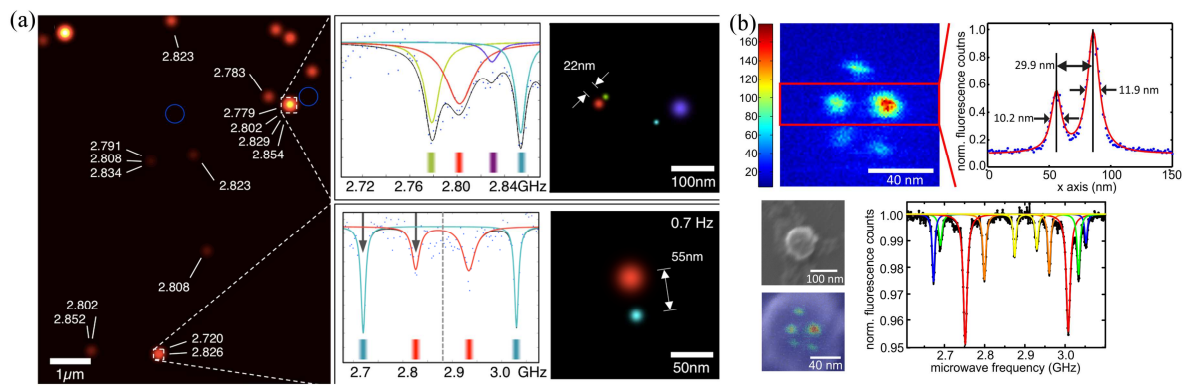

*Figure 11. Super resolution microscopy with NV centres. (a) multicolour imaging using nanodiamonds. Each NV center exhibits different microwave frequency and therefore a different color code. Two examples with separation of 22 nm and 55 nm nanodiamonds are shown. A reconstructed region of nanodiamonds with the blue circles indicating a lack of ESR modulation. Using the ESR spectrum at each site and the deterministic emitter switch microscopy technique, multiemitter sites are reconstructed over a 7 × 9 μm² field of view. The numbers correspond to the resonance frequencies of each NV− in the site. The spectra are ESR signals of a multispectral site. Colored Lorentzian fits correspond to the resonances of each NV− centers in the site, that can subsequently be used for a sub-diffraction limited reconstruction of the NV− centers. reprinted with permission from[142] . (B). (a) Sub diffraction resolution STED image and corresponding vertically binned STED image profile of a diamond particle with ~100 nm diameter showing five isolated NVcenters (red curve: Lorentzian fit). SEM image of the same nanodiamond and overlay of the STED image and the SEM image illustrating the relative dimensions. Confocally recorded ESR spectrum of the same nanodiamond showing five distinct ESR line pairs corresponding to the five NV centers (each line pair fitted with a double Lorentzian in a separate color). Reprinted with permission from [143]*

## 4.4 Raman Lasers

The last application to be discussed is employment of diamond in Raman lasers. The thermal conductivity of diamond is several orders of magnitude higher than other crystals, and combined with its good optical quality, low absorption, loss low coefficient for thermal expansion, diamond has been employed to build small Raman lasers with high average power. Raman lasers are important due to their ability to provide gain in a wide range of the optical spectrum. Indeed, several high power diamond Raman lasers (24.5 W of 1st Stokes power) were demonstrated and very high average powers (>100W) are soon to be realized. Using the first diamond Raman line (energy shift of 1332 cm$^{-1}$) pumped using a 532 nm excitation, lasers at the yellow spectral range (~ 573 nm), which are important for biomedical applications, have been realized. In addition, there is a great interest in realizing lasers in the 'eye-safe' wavelength region in the vicinity of 1.5 μm. at these wavelengths light is strongly absorbed in the eye prior to focusing on the retina, enabling much higher intensities to be safely used in scenarios where inadvertent exposure is considered likely. So far, laser designs were mainly based on erbium doped laser materials or optical parametric oscillators (OPOs), which have limitations in terms of their performance and practicality. Diamond offers an advantage in this regard, as diamond's large Raman shift of 1332 cm$^{-1}$ allows the eye-safe region to be reached from 1.064 μm (a very common pump laser wavelength) in 2 shifts instead of 3, which reduces the complexity of mirror coatings and increases the efficiency. Using a small diamond crystal, 2$^{nd}$ order Stokes conversionto 1.485 μm was demonstrated with conversion efficiencies exceeding 70%[145]. Using an external-cavity, and diamond as a Raman crystal, output powers of up to 14.5 W were achieved at wavelengths of 1240 nm and 1485 nm using a using a 1064 nm pump laser.

Recently, picosecond Raman lasers were realized using diamond. Picosecond lasers are important for many nonlinear optical applications that require non-standard wavelengths. For picosecond operation where the pump laser pulse duration is comparable to, or shorter than the Raman transition dephasing time, the system operates in the transient regime, in contrast to the steady-state regime for CW and nanosecond operation. The Raman gain in the transient regime is smaller for a given peak intensity than in the steady-state regime. To achieve the picosecond diamond laser, a synchronous pumping scheme in a ring cavity configuration was employed. This architecture allowed a selectable method for uni-directional forwards and backwards operation, and efficient 2$^{nd}$ Stokes generation in a singly resonant cavity. Up to 1.0 W output power was obtained in the eye-safe region at 1485 nm, corresponding to 21% overall conversion efficiency[146].

As single crystal diamonds can be grown larger, with minimized birefringence and photonic elements made out of diamond are becoming available - the lasing thresholds will continue to shrink, improving the commercial appeal for diamond Raman lasers. The technology is now looking very promising for compact high power beam converters. It was recently shown that diamond enables brightness conversion of laser beams, in addition to a wavelength shift[147]. The concept underpinning this is well known and established in the 60-70s in gases, however, it is much more challenging for solid-state converters. In this current work, a high power Nd laser that had very poor beam quality (M$^2$~4) was combined with a diamond Raman laser to convert to the eye-safe with a simultaneous enhancement in beam brightness of 50%. Finally, the longest wavelength crystal Raman laser using diamond was demonstrated – enabling tunable output in the mid-infrared from 3.4-3.8 microns. So we are now seeing benefits from diamond fs long wavelength transmission.

# 5  Conclusions and outlook

Despite the challenging nanofabrication of diamond, the progress in diamond nanophotonics has been extremely rapid. In less than a decade, scientists and engineers managed to transform basic polycrystalline devices into high quality photonic crystal cavities suitable for enhancing and studying light matter interactions. New applications are constantly emerging, a very recent one being a realization of a nonlinear photonics platform from diamond, enabled due to the sophisticated cavity fabrication[120]. The time is ripe, therefore, for more detailed and dedicated advanced quantum optical experiments with diamond. These should include proper integrated photonic networks, full exploration of weak coupling and demonstration of strong coupling between a single emitter and a cavity field and multi qubit entanglement. Finally, a deterministic placement of an emitter within the cavity field should be routinely established, either through curving a cavity around a predefined emitter, or a post processing using deterministic chemical or physical doping. The enormous progress toward entangling distant NV centers, strong coupling between the NV center and a superconducting flux qubit[148, 149], improvements in quantum error corrections[150, 151] in conjunction with better optical resonators may (should) result in the demonstration of a solid state, quantum computer in the near future. In the meantime, applications such as quantum key distribution (QKD) (that in part was slowed down by lack of interest from the industry) should not be forgotten. Diamond offers narrowband, ultra bright, single photon emitters in the near infrared that shall be revisited for QKD[53], considering the potential efficiency increase by incorporation into a nanophotonic device.

Novel nanofabrication techniques for diamond, in addition to the fabrication methods introduced in this review, are emerging and will be a focus of research in coming years. For instance, a bottom up approach for diamond photonics was recently realized[152]. In this method, an array of optically active nanodiamonds are grown on a membrane that is subsequently etched. This can yield either standalone nanodiamonds with uniform size or microdisk cavities or waveguides with various shapes. In an alternative fabrication method, it was shown that diamond single crystals that are exposed to 266 nm laser pulses at moderate powers of half of the ablation threshold exhibited etch patterns of hundred nanometers[153]. Furthermore, the patterns were polarization dependent; When the polarization is parallel to [110] (E∥[110]), the pattern consists of faceted ridges oriented perpendicular to the polarization, while if the alignment is along the cubic direction E∥[100], the pattern consists of a cross hatched or grid-like structure oriented within a few degrees of [110]. Another interesting nanopatterning of diamond was realized using electron beam[154] under water vapour. At these conditions, subtractive 3D printing of structures on inclined planes of diamond was realized. Yet, the etch rates in these techniques are extremely slow. Although the processes don't damage the diamond crystal and the optical properties are maintained, at the near future these fabrication avenue is more suitable for prototype devices.

The breadth and the possibilities of various color centers in diamond can be tailored for a diverse range of applications. As such, we have witnessed the discovery of an additional defect with spin dependent fluorescence[88], integration of lanthanides into diamond[89] and a revival of the silicon vacancy center. Furthermore, realization of electrically triggered single emitters[155, 156] may pave the way to scalability and practical, optoelectronic diamond devices. Recent progress in modulating the emission from the negatively charged state is a pivotal advance toward electrical control on single defects in diamond[157, 158] and integration of electrically triggered emitters with photonic resonators. However, more attention should be devoted to behaviour of color centers near interfaces. In particular, stability of emitters in close proximity to surfaces and effect of different surface terminations must be studied in more details[159-161], and for other defects than the NV center. Since many nanophotonic devices require thin (several hundred nanometres thick) membranes, a robust methodology of reliably engineering color centers with close to bulk diamond properties in these thin membranes must be developed. Finally, the exact symmetry and defect structure of many ultra-bright emitters is still unknown, while single emitters in the enviable 1550 nm range are yet to be uncovered. The latest should be mitigated, as single photon detectors in this range becoming available.

The life sciences will certainly benefit from further development of fluorescent nanodiamonds[5, 162]. Yet, their use must extend beyond their current role as a traditional biomarkers or a drug delivery carrier. Future paths may include correlative imaging using single digit nanodiamond or biological processes that will take advantage of the nanodiamond shape. Finally, the combination of high sensitivity temperature[163, 164] and magnetic field sensing should promote the nanodiamonds' use in life sciences and therapeutics[165].

The breakthroughs in diamond nanophotonics were enabled, partly, because of the continuous improvement in the available material. Diamond samples with various impurity concentrations, different synthesis methods (e.g. detonation, CVD and HPHT) and orientations are becoming commercially available. High quality (111) oriented diamond has been engineered recently and yielded enhanced alignment of NVs in photonic structures[38]. To maintain the progress, engineers have to work with scientists and vise versa, to design materials of choice tailored to applications. Through characterization of the device performance, one can learn a lot about the fundamental material properties that will be then harnessed to grow better diamond samples. In the light of these recent advances, diamond has revealed its potential to become the material of choice for commercialization of quantum nanophotonics technologies.

**Acknowledgments**
The authors thank Rich Mildren for discussion of Raman lasers. Igor Aharonovich is the recipient of an Australian Research Council Discovery Early Career Research Award (Project Number DE130100592).